\documentclass[conference]{IEEEtran}
\IEEEoverridecommandlockouts
\usepackage{hyperref}
\usepackage{comment}

\usepackage{cite}
\usepackage{amsmath,amssymb,amsfonts}
\usepackage{algorithmic}
\usepackage{graphicx}
\usepackage{textcomp}
\usepackage{xcolor}

\usepackage{algorithm}
\usepackage{array}
\usepackage{booktabs}
\usepackage{caption}
\usepackage{float}
\usepackage{comment}
\usepackage{multirow}
\usepackage{amsthm}

\newtheorem{theorem}{Theorem}

\def\BibTeX{{\rm B\kern-.05em{\sc i\kern-.025em b}\kern-.08em
    T\kern-.1667em\lower.7ex\hbox{E}\kern-.125emX}}
\begin{document}


\title{Crowdsourced Homophily Ties Based Graph Annotation Via Large Language Model\\

}

\author{\IEEEauthorblockN{Yu Bu}
\IEEEauthorblockA{\textit{Dept. of Computing} \\
\textit{The Hong Kong Polytechnic University}\\
HKSAR\\
uuuyu.bu@connect.polyu.hk}
\and
\IEEEauthorblockN{ Yulin Zhu}
\IEEEauthorblockA{\textit{Dept. of Computing} \\
\textit{Hong Kong Chu Hai College}\\
HKSAR\\
ylzhu@chuhai.edu.hk}
\and
\IEEEauthorblockN{Kai Zhou}
\IEEEauthorblockA{\textit{Dept. of Computing} \\
\textit{The Hong Kong Polytechnic University}\\
HKSAR\\
kaizhou@polyu.edu.hk}
}

\maketitle

\begin{abstract}
Accurate graph annotation typically requires substantial labeled data, which is often challenging and resource-intensive to obtain. In this paper, we present \underline{C}rowd\underline{s}ourced Homophily Ties Based Graph \underline{A}nnotation via Large Language Model (CSA-LLM), a novel approach that combines the strengths of crowdsourced annotations with the capabilities of large language models (LLMs) to enhance the graph annotation process. CSA-LLM harnesses the structural context of graph data by integrating information from 1-hop and 2-hop neighbors. By emphasizing homophily ties—key connections that signify similarity within the graph—CSA-LLM significantly improves the accuracy of annotations. Experimental results demonstrate that this method enhances the performance of Graph Neural Networks (GNNs) by delivering more precise and reliable annotations. 
\end{abstract}


\section{Introduction}


Graph Neural Networks (GNNs) have established themselves as a powerful tool for machine learning on graph-structured data, delivering exceptional performance across a variety of domains\cite{ref1}, including social network analysis\cite{ref2}, recommendation systems\cite{ref3}, and molecular biology\cite{ref4}. However, a significant barrier to their widespread adoption is the necessity for large amounts of labeled data, which is often difficult, costly, and time-consuming to collect\cite{ref5}. To mitigate this challenge, recent research has explored label-free graph annotation methods that leverage Large Language Models (LLMs)\cite{ref6}. These methods aim to reduce reliance on labeled datasets by using LLMs to infer labels based on contextual information\cite{ref7,ref8,ref9,ref10,ref11,ref12,ref13,ref14,ref15}.

\begin{figure}[htb]

\begin{minipage}[b]{1.0\linewidth}
  \centering
  \centerline{\includegraphics[width=8.0cm]{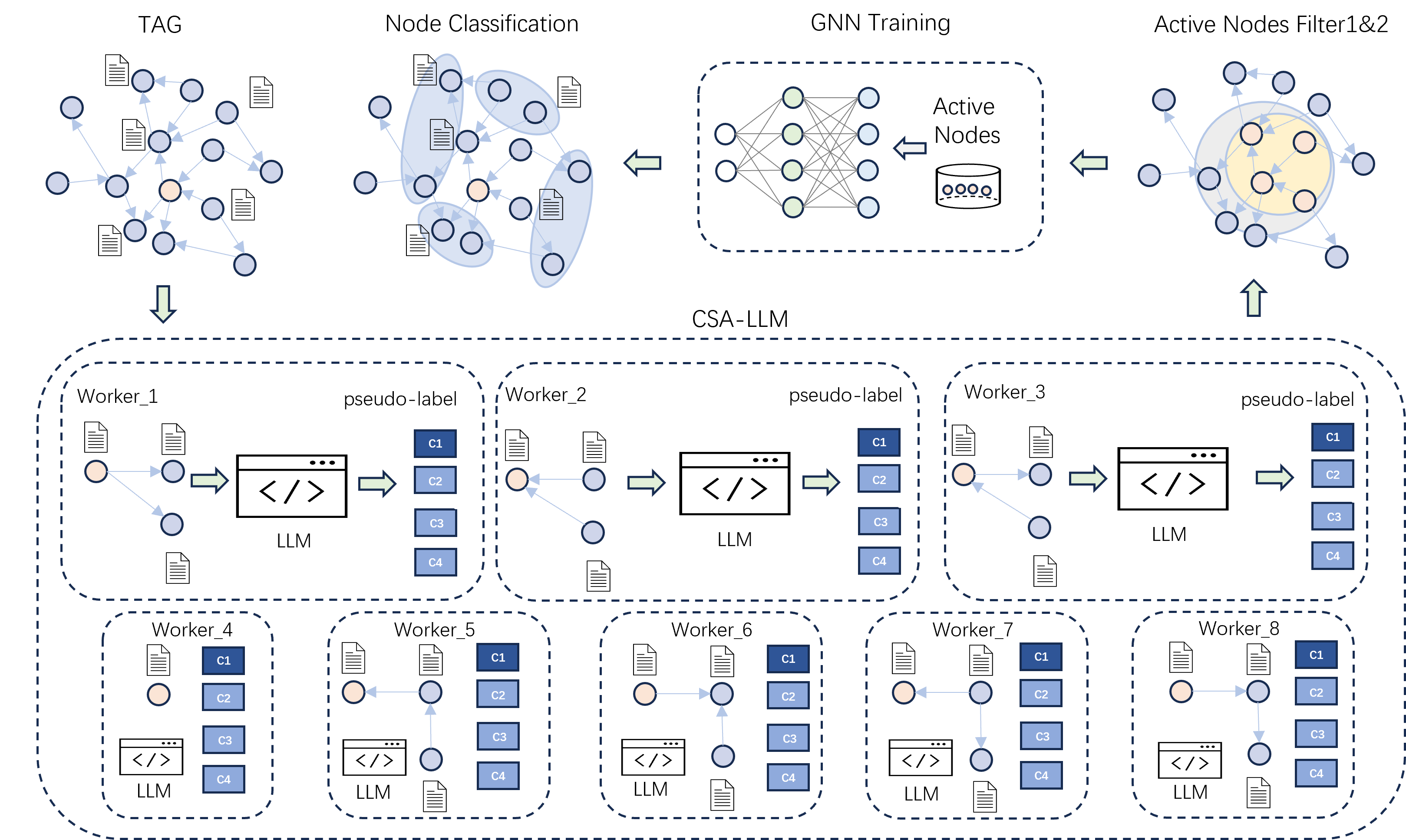}}
\end{minipage}
\hfill
\setlength{\abovecaptionskip}{-0.3cm}
\setlength{\belowcaptionskip}{-0.6cm}
\caption{Crowdsourced Annotation via LLM for GNN.}
\label{fig:framework}

\end{figure}

Despite the potential of LLM-based annotation, some studies have suggested that structural information within graphs has limited impact on improving the accuracy of GNNs when annotated using LLMs \cite{ref6}. These findings raise questions about the effectiveness of incorporating structural context in label-free graph annotations\cite{ref16,ref17,ref18,ref19}. However, our research takes a different approach by leveraging crowdsourced subgraph learning to demonstrate that LLMs can indeed learn from structural information, thereby enhancing the annotation process and ultimately improving GNN performance.

In this paper, we introduce Crowdsourced Homophily Ties Based Graph Annotation via Large Language Model (CSA-LLM), a novel approach that challenges the notion that structure has minimal influence on annotation accuracy. In Figure \ref{fig:framework}, CSA-LLM integrates comprehensive structural context, considering not only individual nodes but also the relationships between nodes and their direct and indirect connections. By focusing on homophily ties \cite{ref20}—key connections that signify similarity within the graph—our approach enables the LLM to make more informed and accurate annotations.

We validate the effectiveness of CSA-LLM through extensive experiments on the Cora and Citeseer datasets \cite{ref21}, two widely used benchmarks in graph-based learning. These datasets, which represent directed text-attributed graphs, provide an ideal testing ground for our method. By incorporating both the directional and textual attributes of these graphs, CSA-LLM effectively harnesses the rich contextual information inherent in the graph structure, resulting in significantly improved annotation accuracy.

Moreover, we introduce a novel filtering strategy \cite{ref6,ref22,ref23} that further enhances our method's performance by prioritizing nodes that are central or influential within the graph, using metrics such as PageRank, degree, and embedding density. This refinement allows CSA-LLM to focus on the most critical parts of the graph, thereby maximizing the effectiveness of the annotations.

Our contributions are as follows: We propose CSA-LLM, a novel graph annotation method that leverages crowdsourced annotations and LLMs, incorporating comprehensive structural context. We introduce a new filtering strategy based on PageRank, degree, and embedding density to enhance annotation accuracy. We demonstrate the effectiveness of CSA-LLM on the Cora and Citeseer datasets, showing significant improvements over baseline approaches. We provide evidence that LLMs can learn from graph-based structural information, opening new avenues for research in label-free GNNs.

The remainder of this paper is structured as follows: Section II reviews preliminaries in GNN and TAG. Section II details the CSA-LLM methodology, including the integration of homopily ties as LLM queries and the new filtering strategy. Section IV presents the importance of homopily ties in theoretic level, followed by experimental setup and results in Section V. Finally, Section VI concludes the paper and highlight the contribution of proposed method.

\section{Preliminaries}
\label{sec:format}

In this section, we introduce the fundamental concepts and notation utilized in our study, which focuses on directed Text-Attributed Graphs (TAGs). A TAG is represented as \( G_T = (V, A, T, X) \), where \( V = \{v_1, v_2, \ldots, v_n\} \) denotes the set of \( n \) nodes in the graph, with each node \( v_i \) corresponding to a distinct entity. Each node \( v_i \) is paired with raw text attributes \( t_i \), forming the set \( T = \{t_1, t_2, \ldots, t_n\} \), which provides semantic context relevant to the node’s role within the graph. These raw text attributes are further encoded into sentence embeddings \( X = \{x_1, x_2, \ldots, x_n\} \) using SentenceBERT \cite{ref24}, a transformer-based model designed to generate high-quality sentence embeddings suitable for various natural language processing (NLP) tasks. The directed connectivity of the graph is captured by the adjacency matrix \( A \in \{0, 1\}^{n \times n} \), where an entry \( A[i, j] = 1 \) indicates a directed edge from node \( v_i \) to node \( v_j \), and \( A[j, i] = 1 \) indicates a directed edge from node \( v_j \) to node \( v_i \). The adjacency matrix is generally asymmetric, reflecting the directed nature of the connections within the graph.

\section{Proposed Method}
\label{sec:pagestyle}

\begin{figure*}[htb]

\begin{minipage}[b]{0.50\linewidth}
  \centering
  \centerline{\includegraphics[width=8.0cm]{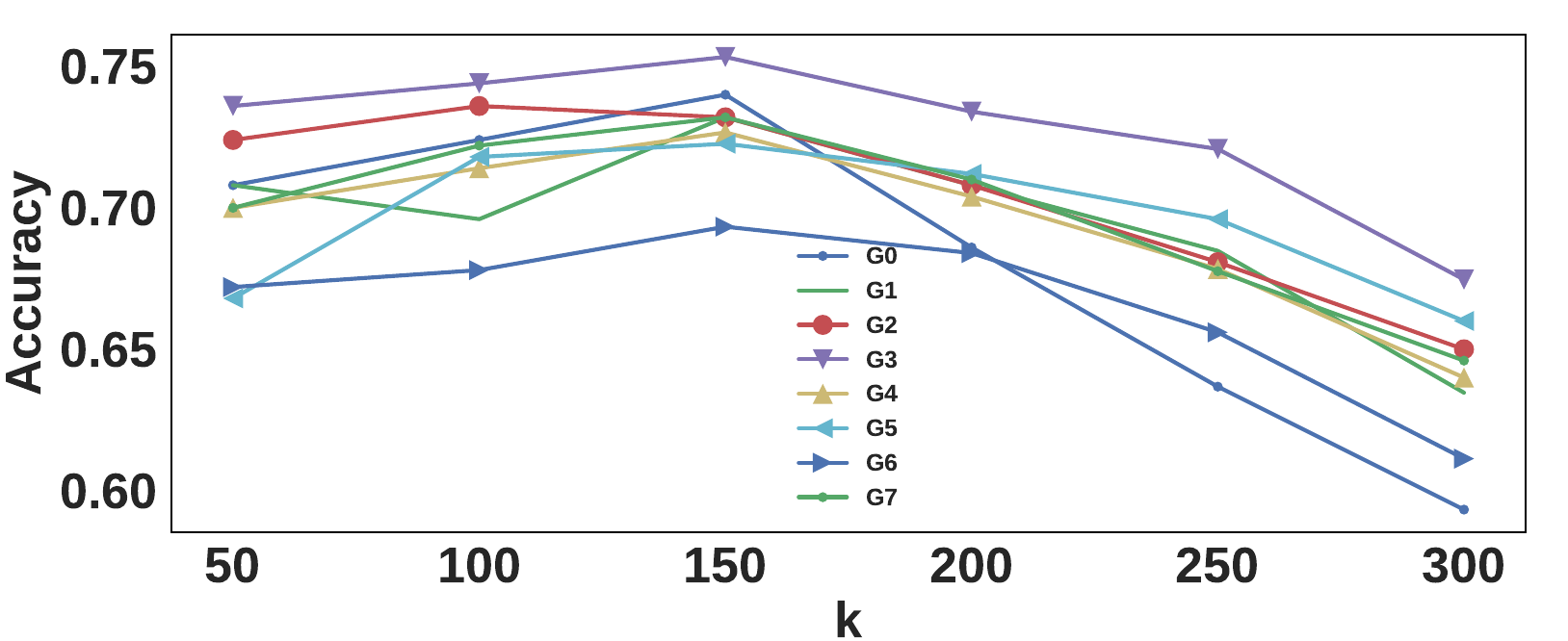}}
\end{minipage}
\hfill
\begin{minipage}[b]{0.50\linewidth}
  \centering
  \centerline{\includegraphics[width=8.0cm]{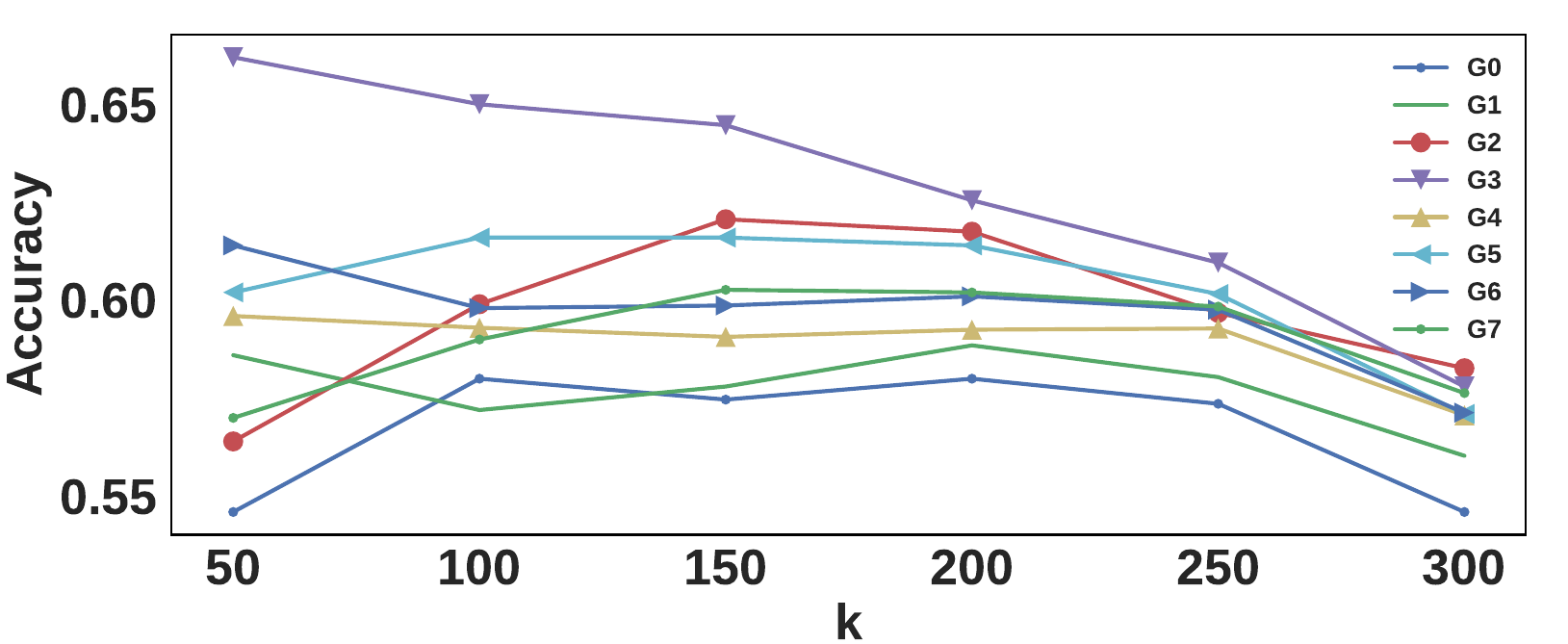}}
\end{minipage}
\setlength{\abovecaptionskip}{-0.3cm}
\setlength{\belowcaptionskip}{-0.7cm}
\caption{Test Accuracy for LLM as Annotator.}
\label{fig:llm_annotator}
\end{figure*}

\subsection{Crowdsourced Annotation via LLM}

In our proposed method, we perform label-free annotation on directed text-attributed graphs by constructing prompts based on the textual information from subgraphs. These subgraphs are centered around a node \( v_i \) and include its neighbors, as well as the neighbors of those neighbors. We consider both 1-hop and 2-hop neighbors, resulting in several possible subgraph configurations that guide the construction of the prompts. Formally, let \( N_i \) represent the set of neighbors within 2-hop of node \( v_i \): 
\[
\begin{array}{cc}
\small
N_i \!=\! \left\{
\begin{aligned}
& N_i^0 \!=\! v_i, \\
& N_i^1 \!=\! \text{pred}(v_i), \\
& N_i^2 \!=\! \text{succ}(v_i), \\
& N_i^3 \!=\! \text{pred}(\text{succ}(v_i)), \\
& N_i^4 \!=\! \text{succ}(\text{pred}(v_i)), \\
& N_i^5 \!=\! \text{pred}(\text{pred}(v_i)), \\
& N_i^6 \!=\! \text{succ}(\text{succ}(v_i))
\end{aligned}
\right\}
&
\mathcal{G}_i^k \!=\! \left\{
\begin{aligned}
& \{ v_i \} , \\
& \{ v_i \} \cup N_i^1, \\
& \{ v_i \} \cup N_i^2, \\
& \{ v_i \} \cup N_i^1 \cup N_i^2, \\
& \{ v_i \} \cup N_i^1 \cup N_i^5, \\
& \{ v_i \} \cup N_i^2 \cup N_i^3, \\
& \{ v_i \} \cup N_i^1 \cup N_i^4, \\
& \{ v_i \} \cup N_i^2 \cup N_i^6
\end{aligned}
\right\}
\end{array}
\]

\begin{table}[h]
\centering
\caption{Full Prompt Example for Zero-Shot \cite {ref6} Annotation on the \textsc{Cora} Dataset}
\label{tab:cora_prompt_example} 
\vspace{-0.2cm}
\renewcommand{\arraystretch}{1.5}
\begin{tabular}{p{0.95\linewidth}}
\specialrule{1.5pt}{0pt}{0pt}
\textbf{Input:} The content of the paper is 
\textbf{text from raw data}
, which is cited by the paper(s) that 
\textbf{text from raw data}. There are following categories: \textbf{list of categories}.
\\
\textbf{Task:} What’s the category of this paper? Provide your 7 best guesses and a confidence number that each is correct (0 to 100) for the following question from the most probable to the least. The sum of all confidence should be 100. For example, [\{"answer": $<$your\_first\_answer$>$, "confidence": $<$confidence\_for\_first\_answer$>$\}, \ldots ] \\
\textbf{Output:} [\{"answer": $<$LLM's\_first\_answer$>$, "confidence": $<$LLM's confidence\_for\_first\_answer$>$\}, \ldots ] \\
\specialrule{1.5pt}{0pt}{0pt}
\end{tabular}
\end{table}

\vspace{-10pt}

\begin{algorithm}
\caption{Active Nodes Filter Strategy}
\label{algorithm_active_nodes}
\begin{algorithmic}[1]
\REQUIRE Graph \(G = (V, A)\), node features \(\mathbf{X}\), annotation confidences \(\mathbf{C}\), hyperparameters \(\gamma, \lambda, \eta\), number of nodes to select \(K\)
\ENSURE Selected nodes for training
\STATE Compute PageRank scores \(\mathbf{P}\); C-Density scores \(\mathbf{D}\), and \(\mathbf{Deg}\)
\STATE Compute first stage filter scores \(S_1(v_i) = \gamma \cdot P(v_i) + \lambda \cdot D(v_i) + (1 - \gamma - \lambda) \cdot \text{Deg}(v_i)\) for all \(v_i \in V\)
\STATE Select top \(K\) nodes based on $S_1(v_i)$

\STATE Compute second stage filter scores \(S_2(v_i) = \text{COE}(v_i) + \text{Confidence}(v_i)\)
\STATE Select top \(K \cdot \eta\) nodes based on $S_2(v_i)$
\STATE \textbf{return} Selected nodes
\end{algorithmic}
\end{algorithm}

\setlength{\textfloatsep}{10pt plus 1.0pt minus 2.0pt}

The textual information for each subgraph \( \mathcal{G}_i^k \) (one homophily tie of $v_i$), where \( k \) indexes the configuration, is aggregated as \( T_{\mathcal{G}_i^k} = \bigcup_{v_j \in \mathcal{G}_i^k} T_{v_j} \), where \( T_{v_j} \) denotes the textual attributes of node \( v_j \). The prompt \( q_i^k \) is then constructed for the LLM by concatenating the textual attributes from all nodes within the subgraph \( \mathcal{G}_i^k \): $q_i^k = \bigg|\bigg|_{v_j \in \mathcal{G}_i^k} T_{v_j}$, where \( \bigg|\bigg| \) represents the concatenation operator, combining the textual information of all nodes within the subgraph into a single input for the LLM. The LLM processes this prompt \( q_i^k \) to generate an annotation \( c_i^k \), which serves as the LLM's output for the graph annotation task. The final annotation for node \( v_i \) is derived from the LLM's responses to the various prompts \( q_i^k \), reflecting a crowdsourced-like aggregation of multiple perspectives within the graph. Table \ref{tab:cora_prompt_example} shows our prompt example for zero-shot annotation. Figure \ref{fig:llm_annotator} presents test accuracy for LLM as annotator from crowdsourced workers.


\begin{figure}[htb]

\begin{minipage}[b]{.48\linewidth}
  \centering
  \centerline{\includegraphics[width=4.0cm]{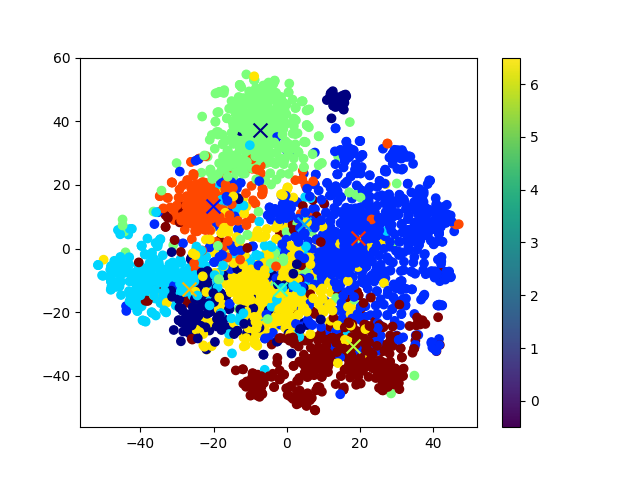}}
\end{minipage}
\hfill
\begin{minipage}[b]{0.48\linewidth}
  \centering
  \centerline{\includegraphics[width=4.0cm]{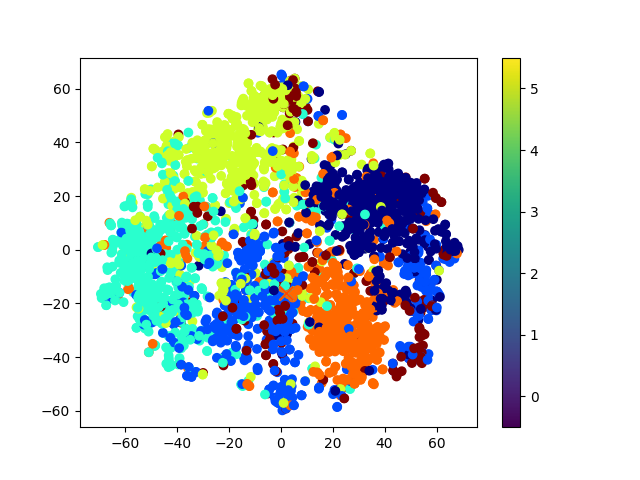}}
\end{minipage}
\setlength{\abovecaptionskip}{-0.1cm}
\setlength{\belowcaptionskip}{-0.3cm}
\caption{Nodes Clustering According to C-Density.}
\label{fig:node_clustering}
\end{figure}

\begin{figure*}[htb]
\begin{minipage}[b]{0.48\linewidth}
  \centering
  \centerline{\includegraphics[width=8.0cm]{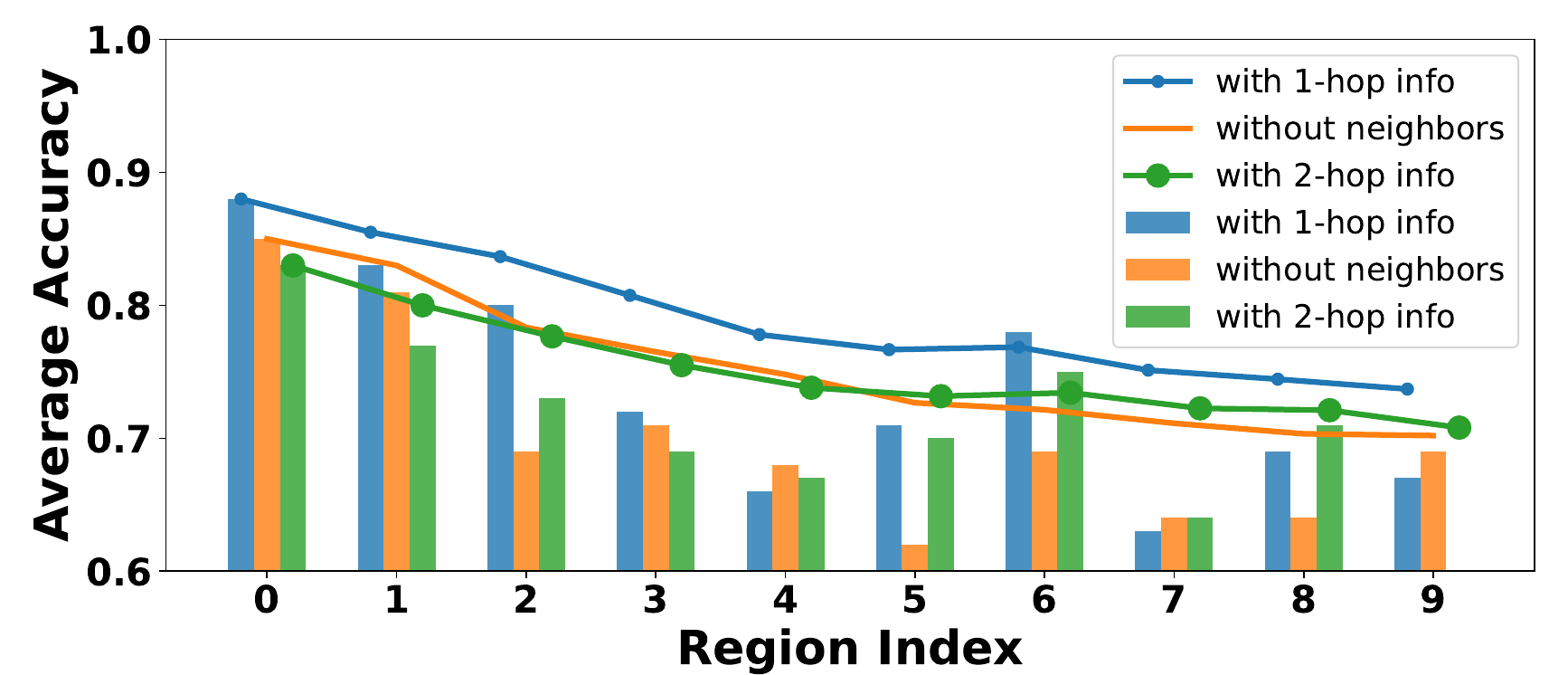}}
\end{minipage}
\hfill
\!
\begin{minipage}[b]{0.48\linewidth}
  \centering
  \centerline{\includegraphics[width=8.0cm]{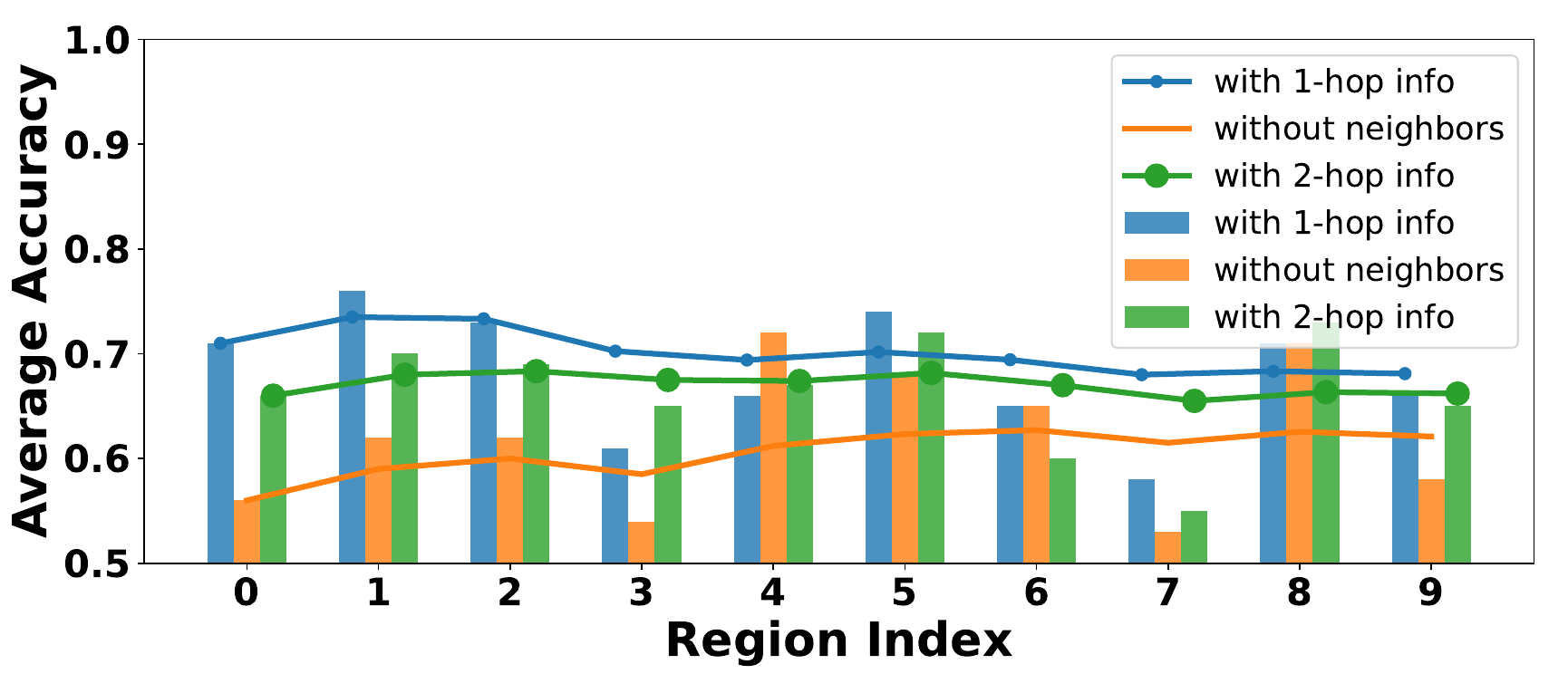}}
\end{minipage}

%
\setlength{\belowcaptionskip}{-0.7cm}
\caption{Test Accuracy According to Distance in C-Density and Confidence Score.}
\!
\label{fig:simple_filter}

\end{figure*}

\begin{figure*}[htb]
\centering
\hspace{-0.7cm}
\begin{minipage}[b]{0.50\linewidth}
  \centering
  \includegraphics[width=8.0cm]{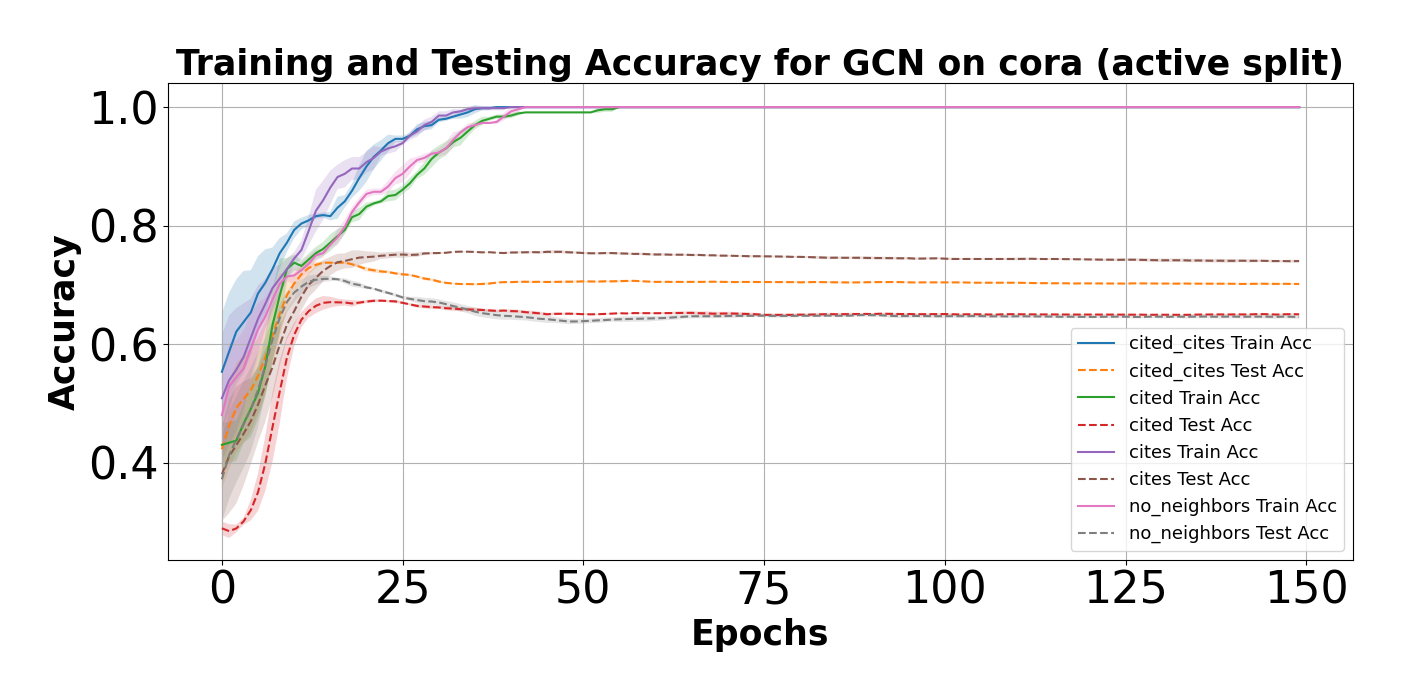}
\end{minipage}
\hfill
\begin{minipage}[b]{0.50\linewidth}
  \centering
  \includegraphics[width=8.0cm]{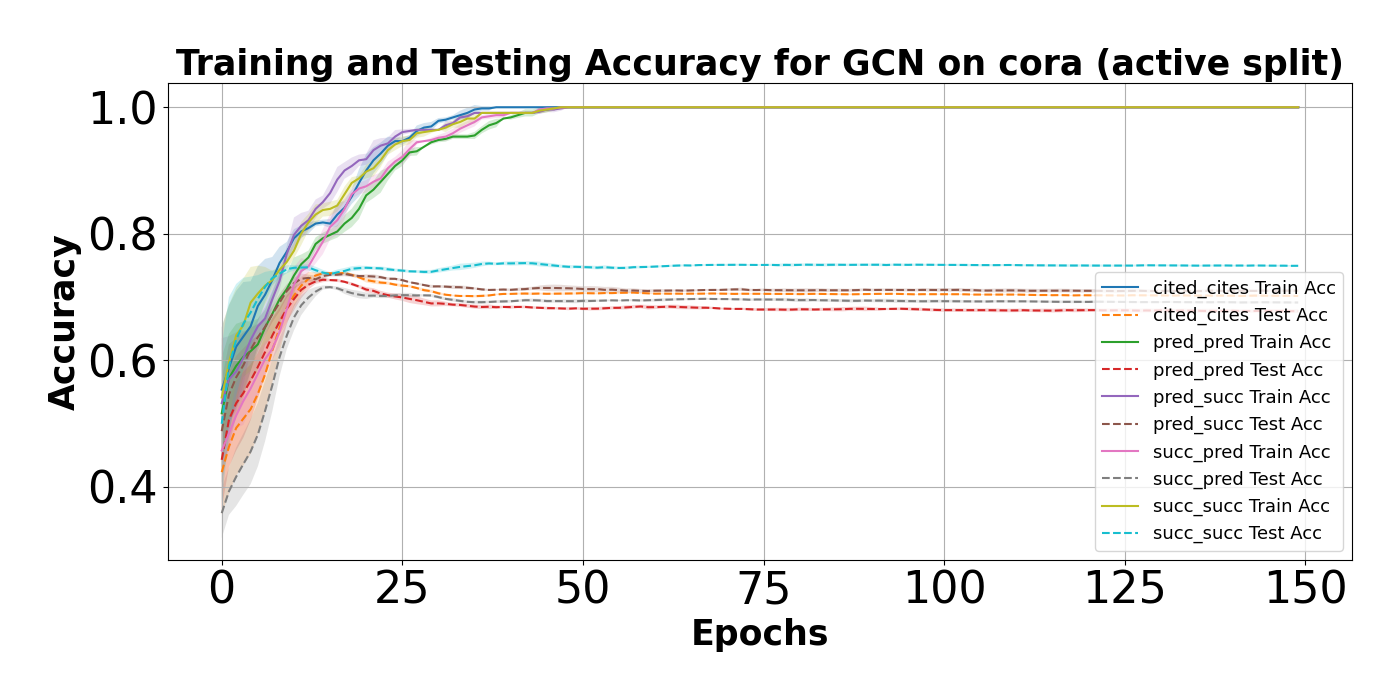}
\end{minipage}

\vspace{-0.3cm}

\hspace{-0.7cm}
\begin{minipage}[b]{0.50\linewidth}
  \centering
  \includegraphics[width=8.0cm]{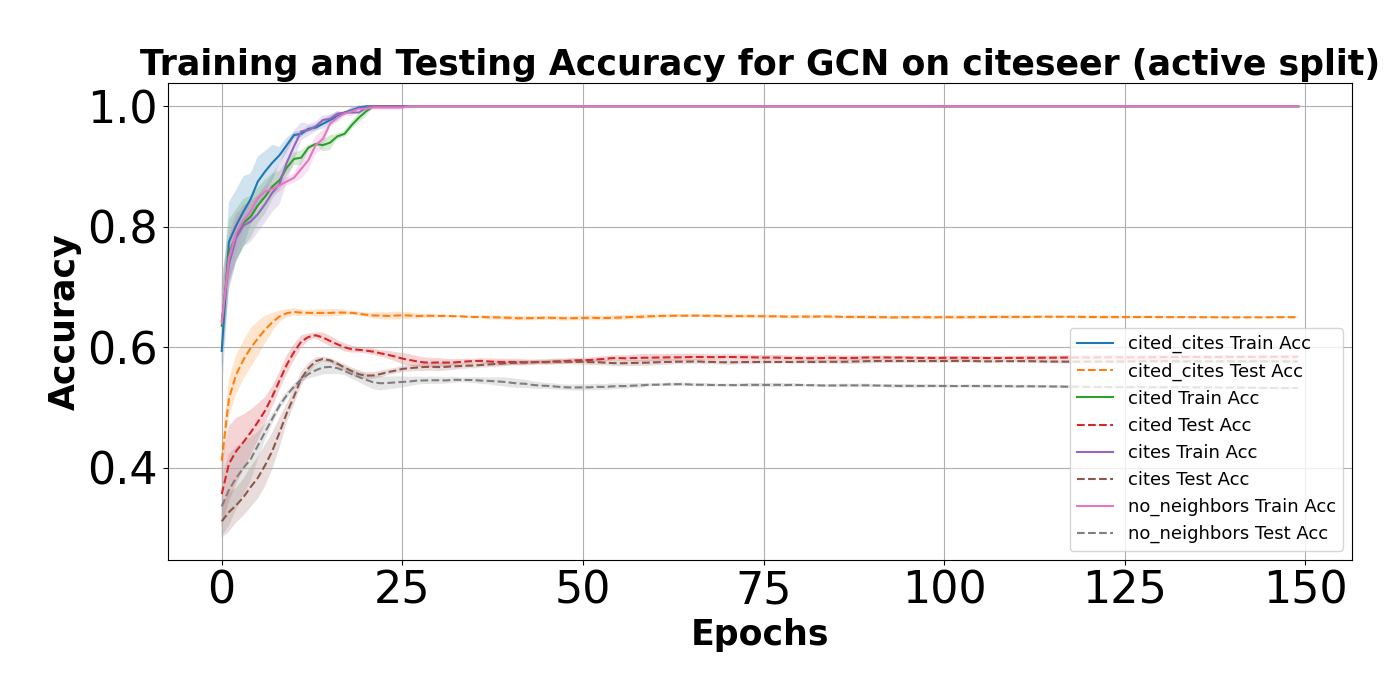}
\end{minipage}
\hfill
\begin{minipage}[b]{0.50\linewidth}
  \centering
  \includegraphics[width=8.0cm]{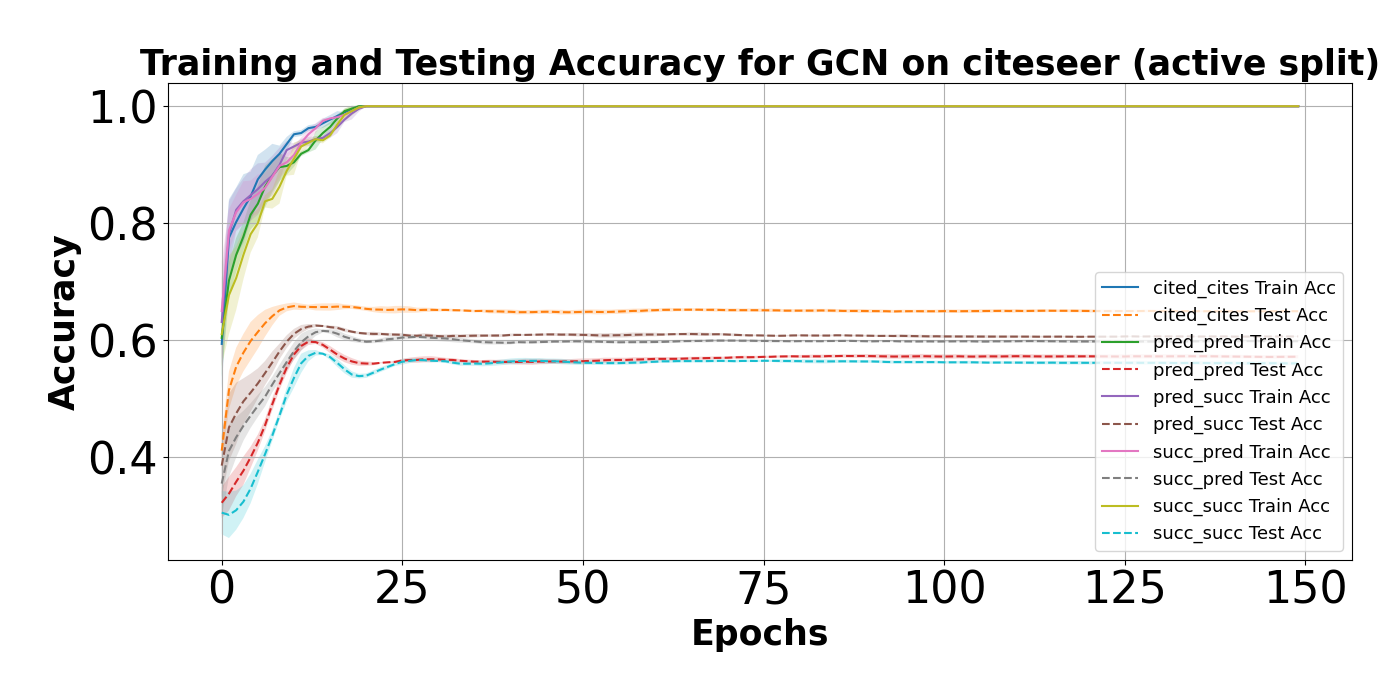}
\end{minipage}
%
\setlength{\abovecaptionskip}{-0.2cm}
\setlength{\belowcaptionskip}{-0.6cm}
\caption{Training and Testing Accuracy for GCN through CSA-LLM.}
\label{fig:res}
\end{figure*}


\subsection{Active Nodes Filter Strategy}

In our proposed method, we implement a filter strategy designed to select the most informative nodes for training the Graph Neural Network (GNN). This strategy combines several metrics, including PageRank score, embedding density, annotation confidence, and node degree, to compute a comprehensive filter score for each node.


The filter score for the first stage of filtering, denoted as \( S_1(v_i) \), is calculated for each node \( v_i \) using the following formula: $S_1(v_i) = \gamma \cdot \text{P}(v_i) + \lambda \cdot \text{C-Density}(v_i) + \theta \cdot \text{Deg}(v_i)$. In this formula, \( \text{P}(v_i) \) represents the PageRank score, which measures the importance of the node within the graph. \( \text{C-Density}(v_i) \) is the cluster-based density score, designed to assess the annotation quality of nodes. This score is calculated as: $\text{C-Density}(v_i) = \frac{1}{1 + \|x_{v_i} - x_{CC_{v_i}}\|}$, where \( x_{v_i} \) denotes the feature vector of node \( v_i \), and \( x_{CC_{v_i}} \) represents the feature vector of the closest cluster center to \( v_i \). The K-means algorithm is used to generate cluster centers from the node embeddings, and the distance to these centers is used as a heuristic for annotation reliability. Figure \ref{fig:node_clustering} shows results of nodes clustering according to C-Density. \( \text{Deg}(v_i) \) denotes the degree of the node, which reflects its connectivity within the graph. The coefficients \( \gamma \), \( \lambda \), and \( \theta \) are hyperparameters that control the contribution of each component to the overall filter score. Nodes are ranked according to their filter scores \( S_1(v_i) \), and the top \( K \) nodes are selected for the first stage of training: $\{v_{i_1}, v_{i_2}, \ldots, v_{i_K}\} = \arg\max_{v_i \in V} S_1(v_i)$.


In the second stage of filtering, we refine the selection by choosing the top \( K \cdot \eta \) nodes based on a score that combines Change of Entropy (COE) and annotation confidence. The score for this stage, denoted as \( S_2(v_i) \), is calculated as: $S_2(v_i) = \text{COE}(v_i) + \text{Confidence}(v_i)$. \( \text{COE}(v_i) \) represents the Change of Entropy for node \( v_i \), computed as: $\text{COE}(v_i) = H(\tilde{y}_{V_{\text{sel}} - \{v_i\}}) - H(\tilde{y}_{V_{\text{sel}}})$. \( H(\cdot) \) is the Shannon entropy function, and \( \tilde{y} \) denotes the annotations generated by the LLMs. Figure \ref{fig:simple_filter} presents the test accuracy simply via C-Density score and confidence score. Algorithm \ref{algorithm_active_nodes} shows the active nodes selection via two stage filters. 





\section{Theoretical Analysis}

In this theorem, we focus on quantifying homophily dominance in multi-hop neighborhoods, building on techniques similar to those used in \cite{ref20}.

\begin{theorem}
Consider a graph \(G\) that does not contain self-loops and has a set of labels \(\mathcal{Y}\) (pseudo-labels generated by LLM). For each node \(v\), assume that the class labels of its neighbors \(\{y_u : u \in N(v)\}\) are conditionally independent given the label \(y_v\) of the node itself. Furthermore, assume that the probability of a neighbor \(u\) having the same label as \(v\) is given by \(P(y_u = y_v \mid y_v) = \alpha\), and the probability of \(u\) having any different label \(y \neq y_v\) is given by \(P(y_u = y \mid y_v) = \frac{1-\alpha}{|\mathcal{Y}|-1}\). Under these conditions: 1. The 1-hop neighborhood \(N_1(v)\) is homophily-dominant in expectation if and only if \(\alpha \geq \frac{1}{|\mathcal{Y}|}\).
2. For any multi-hop neighborhood \(N_h(v)\) with \(h \geq 2\), \(N_h(v)\) will be homophily-dominant in expectation regardless of \(\alpha\), provided \(\alpha > \beta = \frac{1-\alpha}{|\mathcal{Y}|-1}\).

\end{theorem}

\vspace{-0.3cm}

\begin{table}[htbp]
\setcounter{table}{1}
\centering
\caption{Performance Comparison with baselines and different filter strategies on Cora and Citeseer Datasets}
\label{tab:baselines}
\begin{tabular}{lcc}
\specialrule{1.5pt}{0pt}{0pt}
& \textbf{Cora} & \textbf{Citeseer} \\
\specialrule{1.5pt}{0pt}{0pt}
Zero-shot & 67.00 $\pm$ 1.41 & 65.50 $\pm$ 3.53 \\
Zero-shot with hops & 71.75 $\pm$ 0.35 & 62.00 $\pm$ 1.41 \\
\hline
Degree & 68.67 $\pm$ 0.30 & 60.23 $\pm$ 0.54 \\
Degree-W & 69.86 $\pm$ 0.35 & 60.47 $\pm$ 0.49 \\
DA-Degree & 72.86 $\pm$ 0.27 & 60.33 $\pm$ 0.54 \\
PS-Degree-W & 70.92 $\pm$ 0.28 & 62.36 $\pm$ 0.69 \\
DA-Degree-W & 73.01 $\pm$ 0.24 & 61.29 $\pm$ 0.47 \\
Pagerank & 70.31 $\pm$ 0.42 & 61.21 $\pm$ 0.11 \\
Pagerank-W & 71.50 $\pm$ 0.44 & 61.97 $\pm$ 0.19 \\
DA-Pagerank & 74.34 $\pm$ 0.41 & 60.44 $\pm$ 0.40 \\
PS-Pagerank-W & 74.81 $\pm$ 0.37 & 63.27 $\pm$ 0.44 \\
DA-Pagerank-W & 75.62 $\pm$ 0.39 & 61.25 $\pm$ 0.45 \\
\hline
Zero-shot-CSA & \textbf{76.08 $\pm$ 0.50} & \textbf{67.40 $\pm$ 0.57} \\
\specialrule{1.5pt}{0pt}{0pt}
\end{tabular}
\end{table}

\vspace{-0.3cm}

\begin{proof}

Let \(\mathbf{Q}\) represent the transition matrix for label propagation, defined as:
$$\mathbf{Q} = (\alpha - \beta)\mathbf{I} + \beta\mathbf{J},$$ 

where \(\mathbf{I}\) is the identity matrix, \(\mathbf{J}\) is the all-ones matrix, and \(\beta = \frac{1 - \alpha}{|\mathcal{Y}| - 1}\) ensures that each row sums to 1.

\textbf{Eigenvalues of \(\mathbf{Q}\).} Since \(\mathbf{J}\) is a rank-1 matrix with a single nonzero eigenvalue \(|\mathcal{Y}|\), the eigenvalues of \(\mathbf{Q}\) are \(\lambda_1 = 1\) and \(\lambda_2 = \lambda_3 = \dots = \lambda_{|\mathcal{Y}|} = \alpha - \beta\). Using spectral decomposition,
\[
\mathbf{Q}^h = \mathbf{U} \mathbf{\Lambda}^h \mathbf{U}^{-1},
\]
where
\[
\mathbf{\Lambda}^h =
\begin{bmatrix}
1 & 0 & \dots & 0 \\
0 & (\alpha - \beta)^h & \dots & 0 \\
\vdots & \vdots & \ddots & \vdots \\
0 & 0 & \dots & (\alpha - \beta)^h
\end{bmatrix}.
\]
Thus,
\[
\mathbf{Q}^h = (\alpha - \beta)^h \mathbf{I} + \frac{1 - (\alpha - \beta)^h}{|\mathcal{Y}|} \mathbf{J}.
\]

\textbf{Diagonal and Off-Diagonal Elements.} From this, we obtain $[\mathbf{Q}^h]_{i,i} = (\alpha - \beta)^h + \frac{1 - (\alpha - \beta)^h}{|\mathcal{Y}|}, \quad [\mathbf{Q}^h]_{i,j} = \frac{1 - (\alpha - \beta)^h}{|\mathcal{Y}|}, \quad j \neq i.$

\textbf{Proof of Homophily Dominance.} To show that \(N_h(v)\) is homophily-dominant in expectation, we require:
\[
[\mathbf{Q}^h]_{i,i} > [\mathbf{Q}^h]_{i,j}, \quad \forall j \neq i.
\]
Computing the difference, $[\mathbf{Q}^h]_{i,i} - [\mathbf{Q}^h]_{i,j} = (\alpha - \beta)^h + \frac{1 - (\alpha - \beta)^h}{|\mathcal{Y}|} - \frac{1 - (\alpha - \beta)^h}{|\mathcal{Y}|} = (\alpha - \beta)^h.$

Since \(\alpha > \beta\), we have \((\alpha - \beta)^h > 0\), and thus for all \(h \geq 2\),
\[
[\mathbf{Q}^h]_{i,i} > [\mathbf{Q}^h]_{i,j}, \quad \forall j \neq i.
\]
This proves that multi-hop label propagation preserves homophily dominance.

\textbf{Conclusion.} The 1-hop neighborhood \(N_1(v)\) is homophily-dominant if and only if \(\alpha \geq \frac{1}{|\mathcal{Y}|}\). For any \(h \geq 2\), the \(h\)-hop neighborhood \(N_h(v)\) remains homophily-dominant in expectation if \(\alpha > \beta\).

\end{proof}

\section{Evaluation}
\label{sec:typestyle}

We first compare the performance of CSA-LLM against several baseline methods in \cite{ref11} and different filter strategies in \cite{ref6}. The results, presented in Table \ref{tab:baselines}, demonstrate that CSA-LLM consistently outperforms the baselines across both Cora and Citeseer datasets. We further analyze the impact of hyperparameter settings on the test accuracy of the GNN models. Table \ref{tab:hyper} provides a detailed breakdown of the hyperparameters used in our experiments, underscoring the importance of hyperparameter selection in graph-based learning tasks. To assess the learning capability of the GNN models, we track the training and testing accuracy over multiple epochs. The results, illustrated in Figure \ref{fig:res}, show a each worker's accuracy as the model converges. We control the total cost for all crowdsourced workers using ChatGPT 3.5 to around 2 dollars per dataset.

\vspace{-0.2cm}

\begin{table}[htbp]
\centering
\caption{Hyperparameters and Test Accuracy for Cora and Citeseer Datasets}
\label{tab:hyper}
\begin{tabular}{|c|c|c|}
\hline
\textbf{Dataset} & \textbf{Hyperparameters} & \textbf{Test Acc} \\ \hline
\multirow{5}{*}{$\! \textsc{Cora}\quad{\eta=0.15} $} 
& $\gamma = 0.00$, $\lambda = 0.80$ & 75.71 $\pm$ 0.68 \\ \cline{2-3} 
& $\gamma = 0.01$, $\lambda = 0.79$ & 74.96 $\pm$ 0.42 \\ \cline{2-3} 
& $\gamma = 0.02$, $\lambda = 0.78$ & 76.08 $\pm$ 0.50 \\ \cline{2-3} 
& $\gamma = 0.03$, $\lambda = 0.77$ & 74.49 $\pm$ 0.20 \\ \cline{2-3} 
& $\gamma = 0.04$, $\lambda = 0.76$ & 74.99 $\pm$ 0.29 \\ \cline{2-3} 
 \hline
\multirow{5}{*}{$\! \textsc{Citeseer}\quad{\eta=0.20} $} 
& $\gamma = 0.07$, $\lambda = 0.73$ & 65.40 $\pm$ 0.49 \\ \cline{2-3} 
& $\gamma = 0.08$, $\lambda = 0.72$ & 66.28 $\pm$ 0.65 \\ \cline{2-3} 
& $\gamma = 0.09$, $\lambda = 0.71$ & 67.40 $\pm$ 0.57 \\ \cline{2-3} 
& $\gamma = 0.10$, $\lambda = 0.70$ & 66.32 $\pm$ 0.56 \\ \cline{2-3} 
& $\gamma = 0.11$, $\lambda = 0.69$ & 64.01 $\pm$ 0.57 \\ \cline{2-3} 
\hline
\end{tabular}
\end{table}


\vspace{-0.5cm}

\section{Conclusion}
In conclusion, the CSA-LLM approach effectively addresses the challenges of graph annotation by leveraging the combined power of crowdsourced annotations and large language models. Through its focus on homophily ties, CSA-LLM enhances the structural understanding of graph data, leading to more accurate and reliable annotations. This, in turn, significantly improves the performance of Graph Neural Networks. 



\end{document}